\documentclass{iaus}
\usepackage{graphicx}

\title[Effects of Supernova Feedback on Galaxy Formation] 
{Effects of Supernova Feedback on the Formation of Galaxies}

\author[Cecilia Scannapieco et al.]   
{Cecilia Scannapieco$^1$, Patricia B. Tissera$^2$, Simon D.M. White$^1$ 
 \and Volker Springel$^1$}

\affiliation{$^1$ Max-Planck Institute for Astrophysics\\ Karl-Schwarzchild Str. 1, D85748, Garching, Germany \\ 
email: {\tt cecilia@mpa-garching.mpg.de} \\[\affilskip]
$^2$  Instituto de Astronom\'{\i}a y F\'{\i}sica del Espacio\\ Casilla de Correos 67,
Suc. 28, 1428, Buenos Aires, Argentina}

\pubyear{2008}
\volume{254}  
\pagerange{1-6}
\jname{The Galaxy Disk in Cosmological Context}
\editors{J. Andersen, J. Bland-Hawthorn \& B. Nordstr\"om, eds.}
\begin{document}

\maketitle

\begin{abstract}
We study the effects of Supernova (SN) feedback on
the formation of galaxies using hydrodynamical simulations
in a
$\Lambda$CDM cosmology.
We use an extended version of the code GADGET-2
 which includes chemical
enrichment and energy feedback by Type II and
Type Ia SN, metal-dependent cooling and
a multiphase model for the gas component.
We focus on the effects of SN feedback on the
star formation process, galaxy morphology, evolution
of the specific angular momentum and chemical properties.
We find that SN feedback plays a fundamental role
in galaxy evolution, producing a self-regulated cycle
for star formation, preventing the early consumption
of gas and allowing disks to form at late times.
The SN feedback model is able to reproduce the expected dependence
on virial mass, with less massive systems being more strongly
affected. 
\keywords{galaxies: formation, galaxies: evolution, methods: n-body simulations}
\end{abstract}

\firstsection 
\section{Introduction}

Supernova  explosions play a fundamental role in galaxy formation
and evolution. On one side, they are the main source
of heavy elements in the Universe and the presence of
such elements substantially enhances the cooling of
 gas (White \& Frenk 1991). On the other hand, SNe
eject a significant amount of energy into the interstellar medium.
It is believed that SN explosions are responsible of generating
a self-regulated cycle for star formation through the heating
and disruption of cold gas clouds, as well as of triggering
important galactic winds such as those observed (e.g. Martin 2004).
Smaller systems are more strongly affected by SN feedback,
because  their shallower potential
wells  are less efficient in retaining baryons (e.g. White \& Frenk 1991).

Numerical simulations have become an important tool to study
galaxy formation, since  they can track the joint evolution
of dark matter and baryons in the context of a cosmological
model. However, this has shown to be an extremely
complex task, because of the need to cover
a large dynamical range and describe, at the
same time, large-scale processes such as tidal
interactions and mergers and small-scale processes
related to stellar evolution.

One of the main problems that  galaxy formation
simulations have repeteadly found is the inability to reproduce
the  morphologies of disk galaxies observed in the Universe.
This is generally refered to as the angular momentum problem
that arises when baryons transfer most of their
angular momentum to the dark matter components during
interactions and mergers (Navarro \& Benz 1991; Navarro
\& White 1994). As a result, disks are too small and concentrated
with respect to real spirals.
More recent simulations which include prescriptions for
SN feedback have been able to produce more realistic
disks (e.g. Abadi et al. 2003; Robertson et al. 2004; Governato
et al. 2007). These works have pointed out the importance
of SN feedback as a key process to prevent the
loss of angular momentum, regulate the star formation
activity and produce extended, young disk-like components.

In this work, we investigate the effects of SN feedback
on the formation of galaxies, focusing on the
formation of disks. For this purpose, we have
run simulations of a Milky-Way type galaxy
using an extended version of the code {\small GADGET-2}
which includes chemical enrichment and energy feedback by SN.
A summary of the simulation code and the initial conditions
is given in Section~\ref{simus}. In Section~\ref{results}
we investigate the effects of SN feedback on galaxy morphology,
star formation rates,  evolution of specific
angular momentum and chemical properties.
We also investigate the dependence of the results on virial mass.
Finally, in Section~\ref{conclusions} we give our conclusions.

\section{Simulations}\label{simus}

We use the simulation code described in Scannapieco et
al. (2005, 2006). This is an extended version of
the Tree-PM SPH code {\small GADGET-2} (Springel \& Hernquist
2002; Springel 2005), which includes chemical enrichment
and energy feedback by SN, metal-dependent cooling
and a multiphase model for the gas component.
Note that our star formation and feedback model is substantially different from that
of Springel \& Hernquist (2003), but we do include their treatment
of UV background.

We focus on the study of a disk galaxy similar to the Milky Way 
in its cosmological context. For this purpose  we
simulate a system with $z=0$ halo mass of  $\sim 10^{12}$ $h^{-1}$ M$_\odot$
and spin parameter of $\lambda\sim 0.03$,
extracted from a large cosmological simulation and
resimulated with improved resolution.
It was selected to have no major mergers
since $z=1$ in order to give time for a disk to form.
The simulations adopt a $\Lambda$CDM Universe
with the following cosmological parameters: 
$\Omega_\Lambda=0.7$, $\Omega_{\rm m}=0.3$,
$\Omega_{\rm b}=0.04$,  a normalization of the power spectrum of
$\sigma_8=0.9$ and 
$H_0=100\ h$ km s$^{-1}$ Mpc$^{-1}$ with $h=0.7$.
The particle mass is
$1.6\times 10^7$ for dark matter and $2.4\times 10^6$ $h^{-1}$
M$_\odot$  for baryonic particles, and we use  a maximum gravitational 
softening of $0.8\ h^{-1}$ kpc
for gas, dark matter and star particles.
At $z=0$ the halo of our galaxy  contains $\sim 1.2\times 10^5$ dark matter
and $\sim 1.5\times 10^5$  baryonic particles within the virial radius.

In order to investigate the effects of SN feedback on
the formation of galaxies, we compare two simulations
which only differ in the inclusion of the SN energy
feedback model. These simulations are part of the
series analysed in Scannapieco et al. (2008), where
an extensive investigation of the effects of SN feedback on
galaxies and a parameter study is performed.
In this work, we use the no-feedback run NF (run without
including the SN energy feedback model) and the
feedback run E-0.7. We refer the interested reader
to Scannapieco et al. (2008) for details in the
characteristics of these simulations.

\section{Results}\label{results}

\begin{figure}[h]
\begin{center}
 \includegraphics[width=60mm]{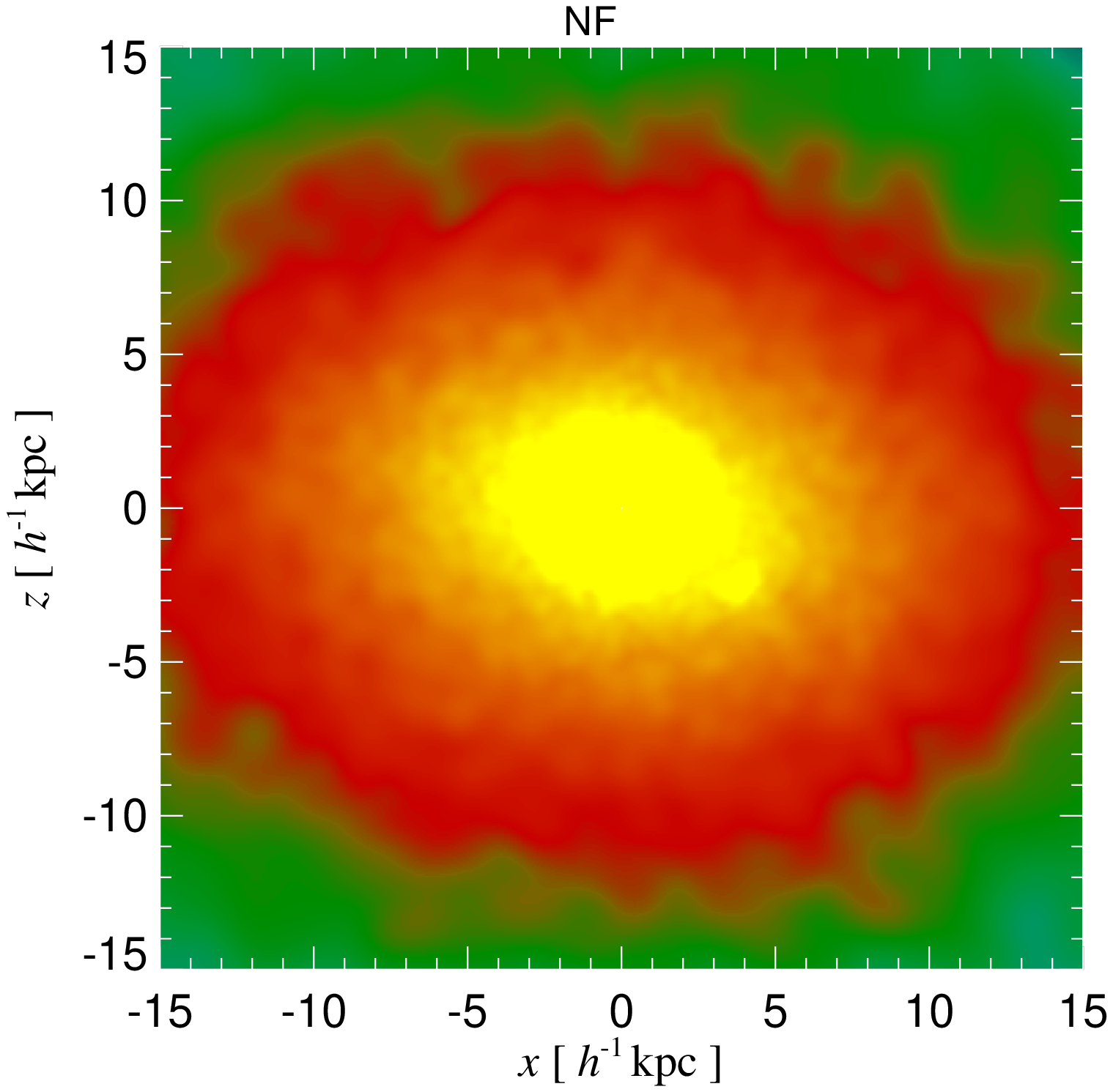} \includegraphics[width=60mm]{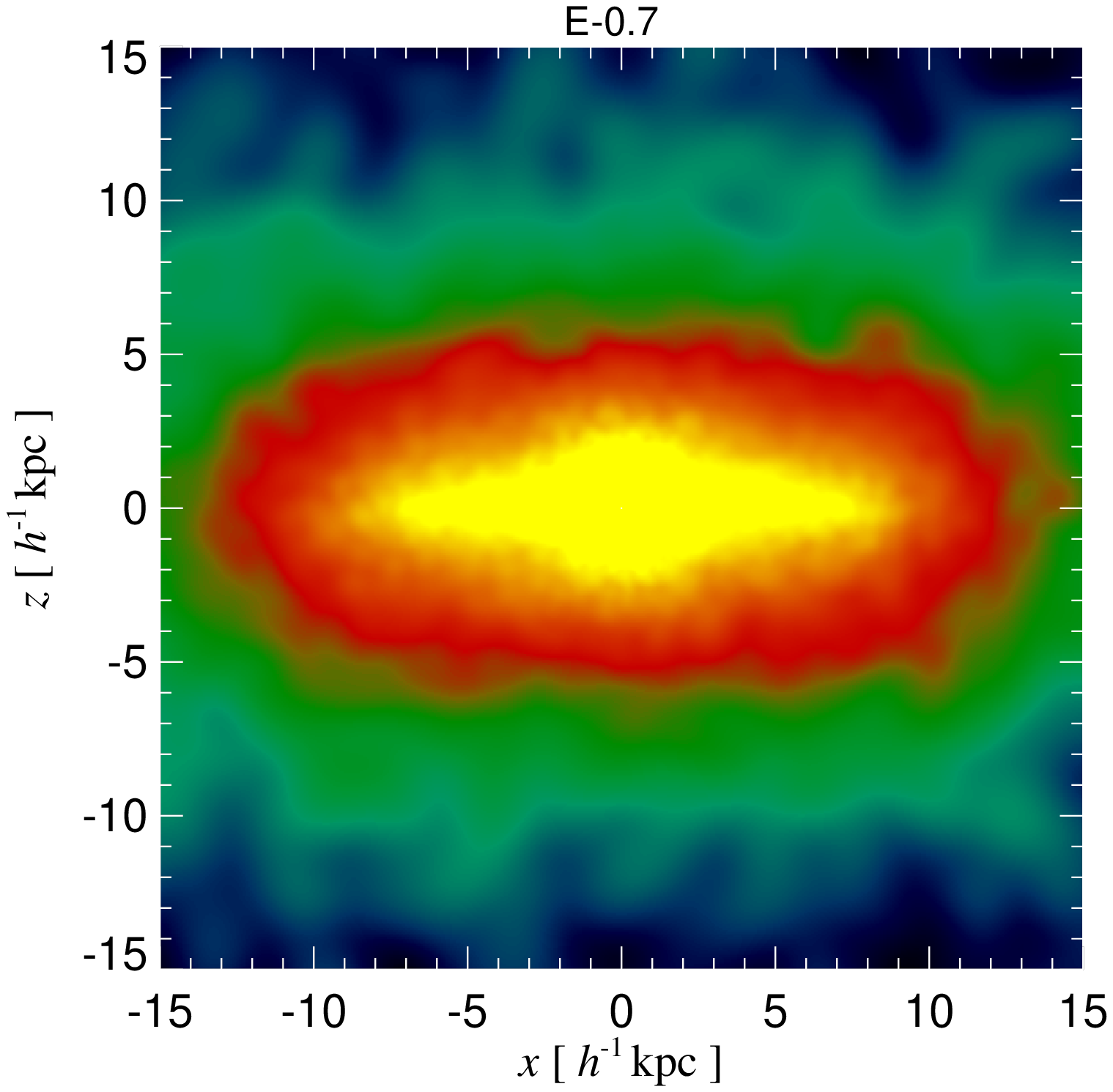} 
 \caption{Edge-on stellar surface density maps for the no-feedback (NF,
left-hand panel)
and feedback (E-0.7, right-hand panel) simulations
at $z=0$.  The colors span 4 orders
of magnitude in projected density, with brighter colors
representing higher densities.  
}
   \label{maps}
\end{center}
\end{figure}

In Fig.~\ref{maps}
we show stellar surface density maps at $z=0$ for the NF and E-0.7
runs. Clearly, SN feedback has an important effect
on the final morphology of the galaxy. If SN feedback
is not included, as we have done in run NF, the stars
define a spheroidal component with no disk.
On the contrary, the inclusion of SN energy feedback
allows the formation of an extended disk component.

\begin{figure}[h]
\begin{center}
 \includegraphics[width=70mm]{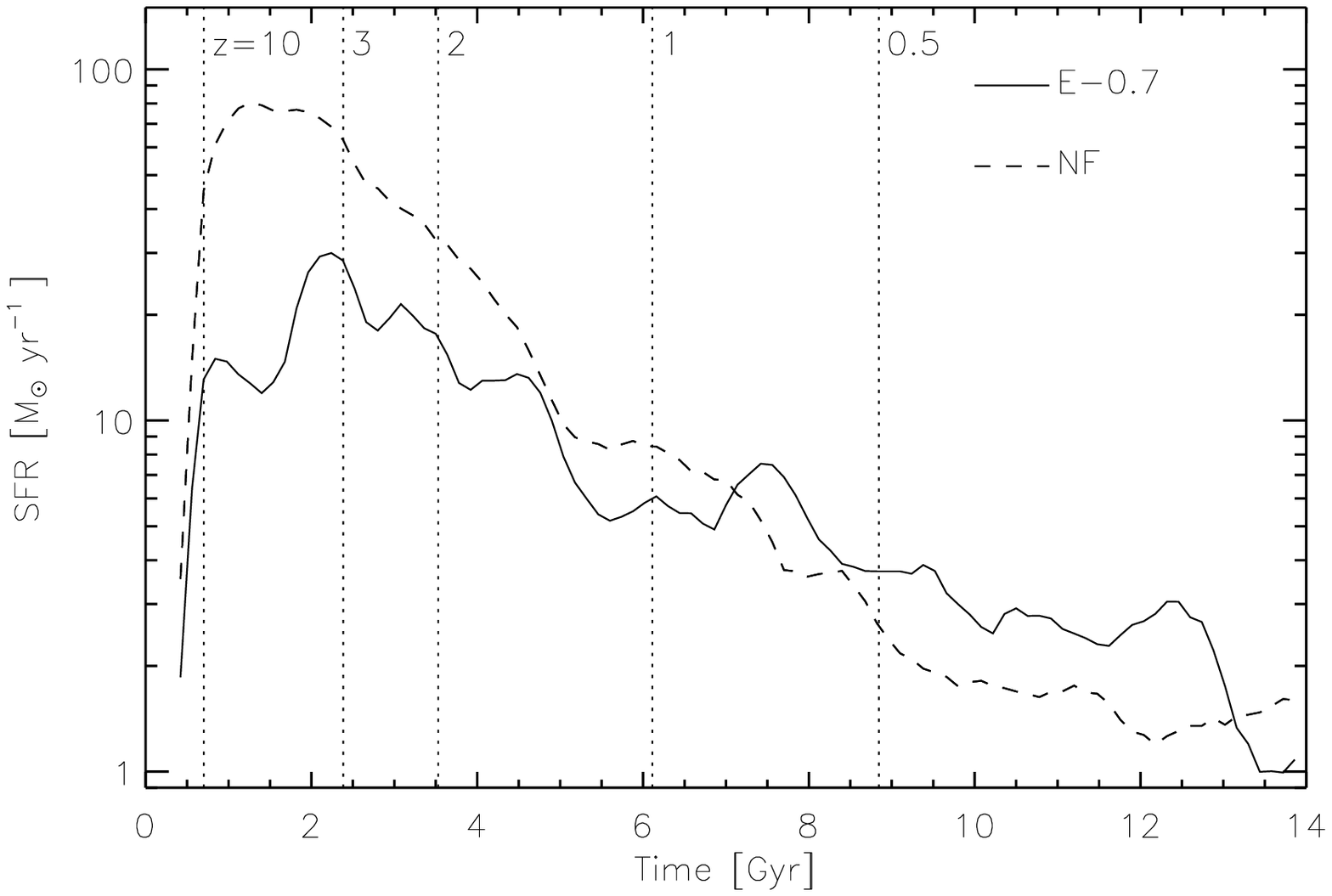}\includegraphics[width=64mm]{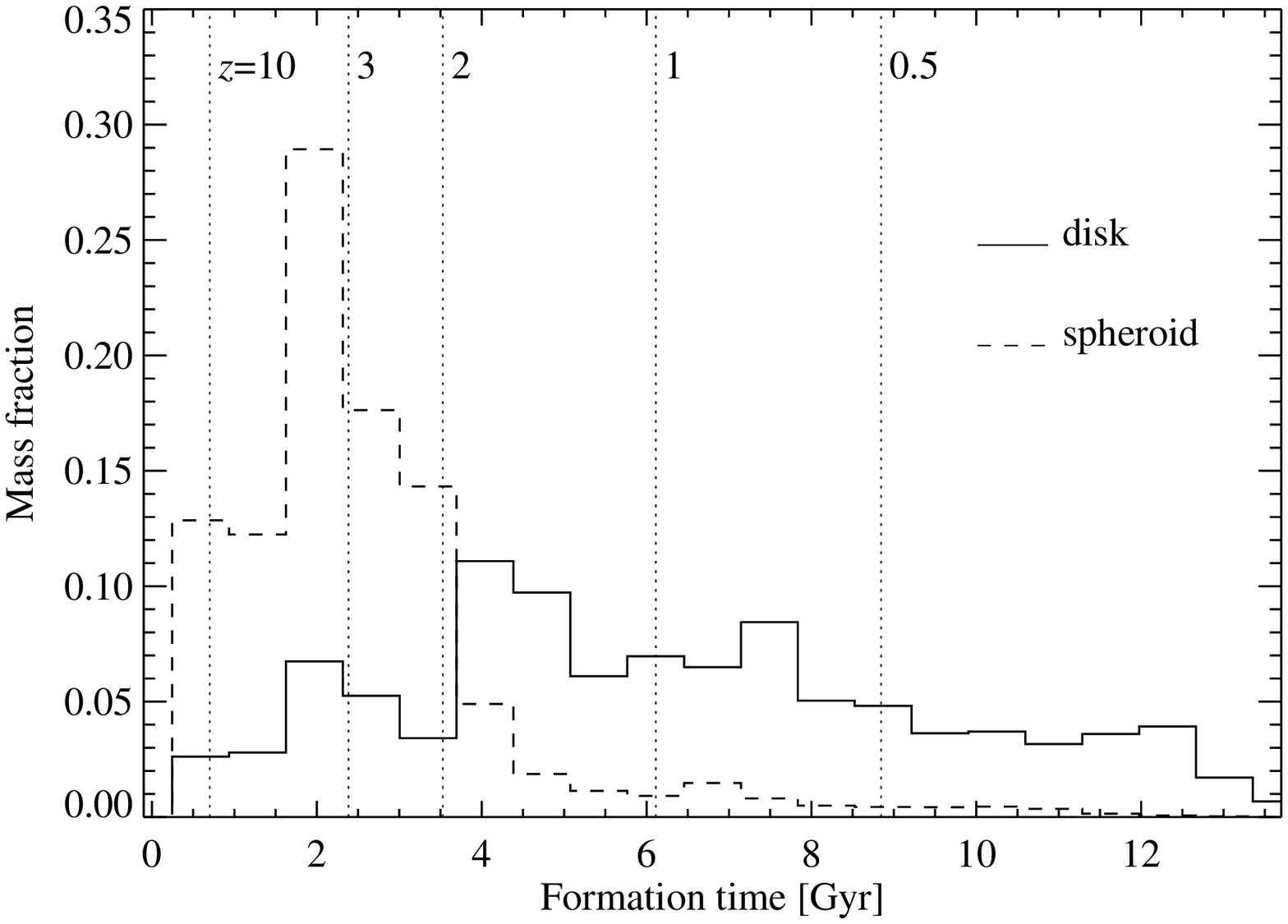} 
 \caption{Left: Star formation rates for the no-feedback (NF)  and feedback (E-0.7) runs.
Right: Mass fraction as a function of formation time for stars
of the disk and spheroidal components in simulation E-0.7. }
   \label{sfr_stellarage}
\end{center}
\end{figure}

The generation of a disk component is
closely related to the star formation process.
In the left-hand panel of Fig.~\ref{sfr_stellarage} we show the star formation rates (SFR) for our
simulations. In the no-feedback case (NF), the gas cools down and
concentrates at the centre of the potential well very early,
producing a strong starburst which feeds the galaxy spheroid. As a result of the early
consumption of gas to form stars, the SFR is  low at
later times. On the contrary, the SFR obtained for
the feedback case is lower at early times, indicating
that SN feedback has contributed to self-regulate the
star formation process. This is the result of the
heating of gas and the generation of galactic winds.
In this case, the amount of gas available for star
formation is larger at recent times and consequently the SFR is higher. In 
the right-hand panel of Fig.~\ref{sfr_stellarage} we show 
the mass fraction as a function of formation time for stars
of the disk and spheroidal components in our feedback
simulation (see Scannapieco et al. 2008 for the method
used to segregate stars into disk and spheroid).
From this plot it is clear that  star formation
at recent times ($z\lesssim 1$) significantly contributes
to the formation of the disk component, while stars
formed at early times contribute mainly to
the spheroid. In this simulation, 
$\sim 50$ per cent of the mass of the disk forms since $z=1$.
Note that in the no-feedback case, only
a few per cent of the final stellar mass of the galaxy
is formed since $z=1$.

Our simulation E-0.7 has produced a galaxy with
an extended disk component. By using the segregation
of stars into disk and spheroid mentioned above, we can calculate
the masses of the different components, as well as
characteristic scales.
The disk of the simulated galaxy has a mass
of $3.3\times 10^{10}\ h^{-1}\ M_\odot$, a
half-mass radius of $5.7\ h^{-1}$ kpc, a half-mass
height of $0.5\ h^{-1}$ kpc, and a half-mass formation
time of $6.3$ Gyr. The spheroid mass and half-mass
formation time are $4.1\times 10^{10}\ h^{-1}\ M_\odot$ 
and $2.5$ Gyr, respectively. It is clear that
the characteristic half-mass times are very different
in the two cases, the disk component being  formed
by younger stars.

\begin{figure}[h]
\begin{center}
 \includegraphics[width=65mm]{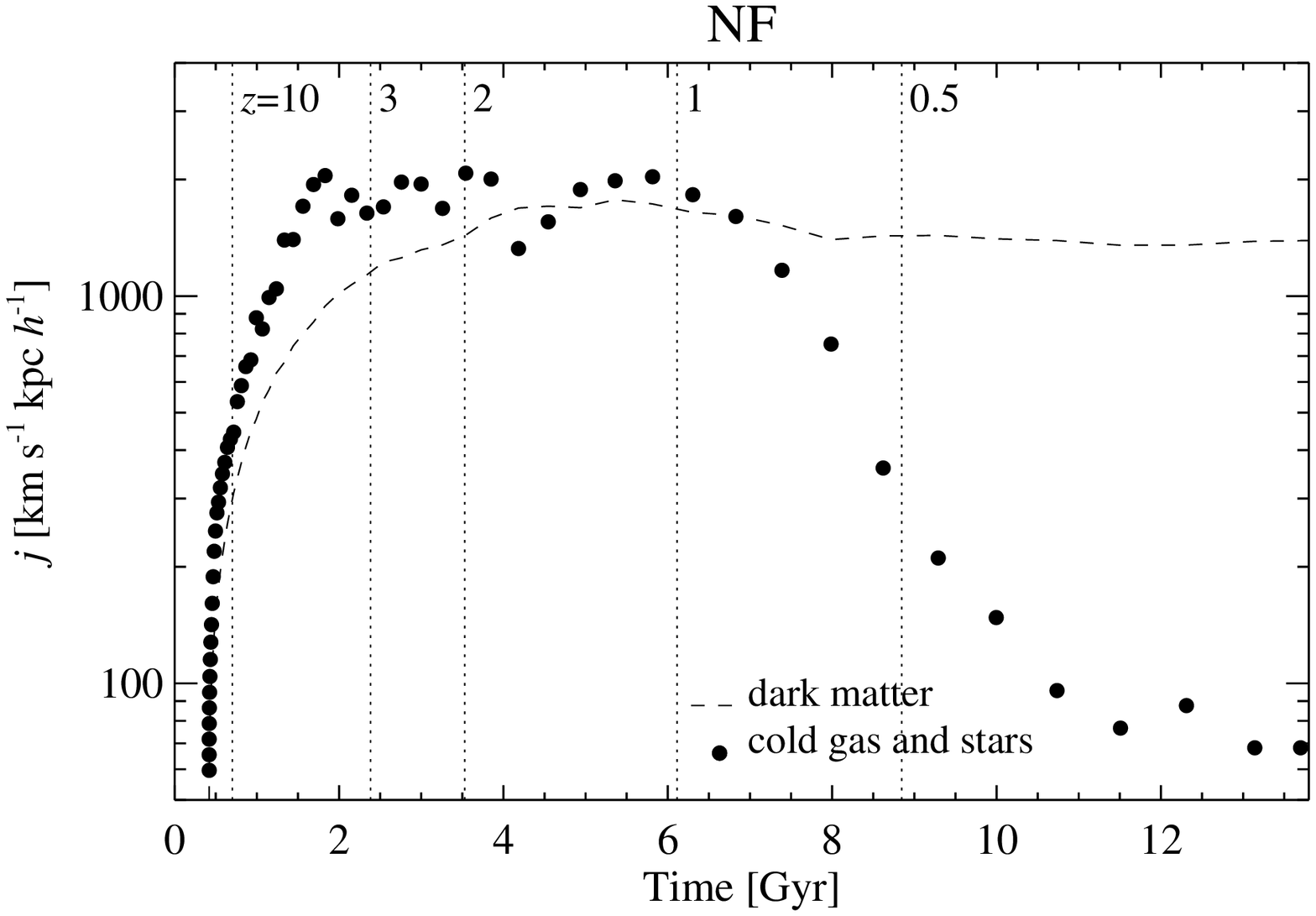} \includegraphics[width=65mm]{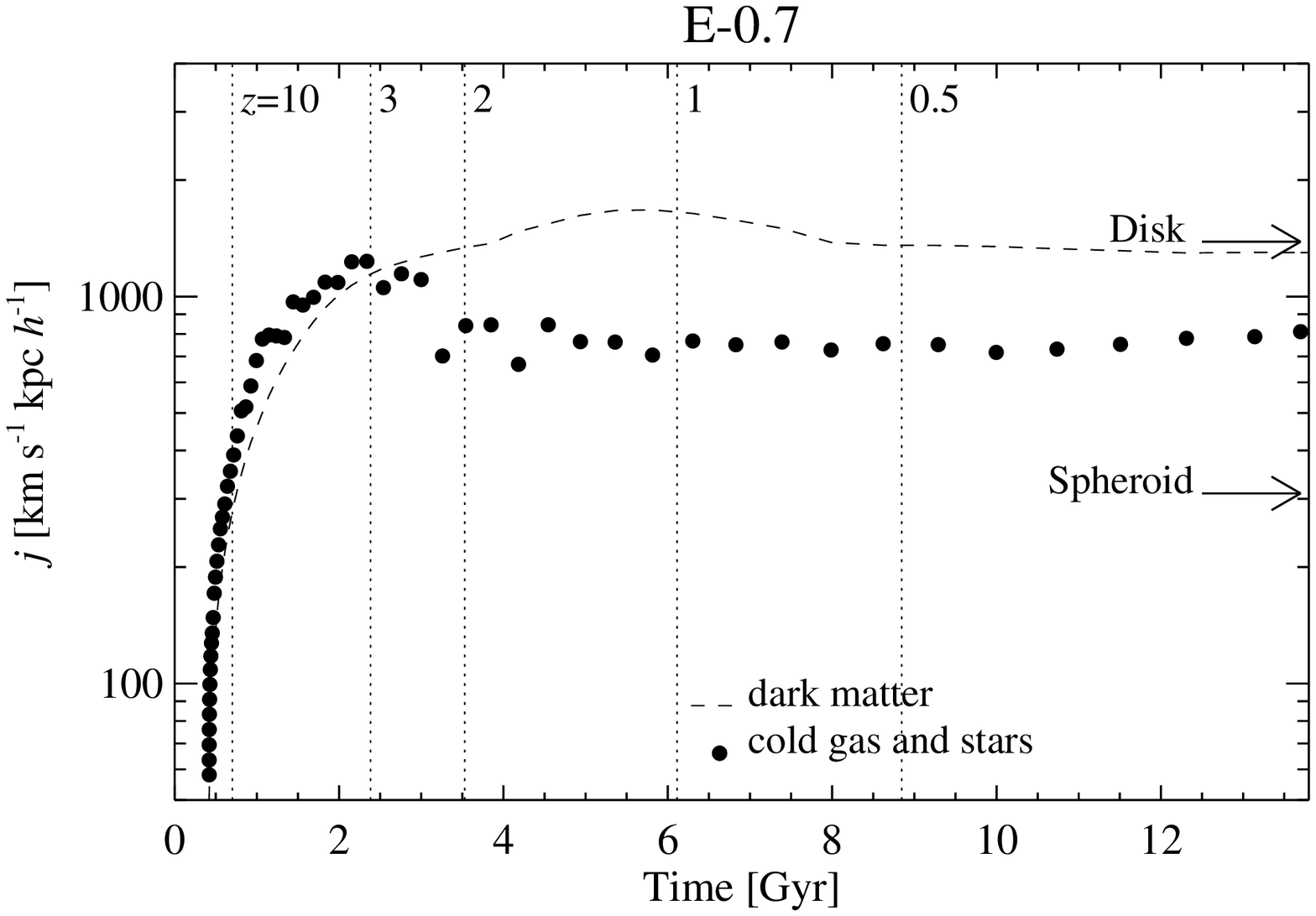} 
 \caption{Dashed lines show the specific angular momentum as a function of time 
for the dark matter that, at $z=0$, lies within the virial radius of
the system for  NF (left panel) and E-0.7 (right panel).
We also show with dots the specific angular momentum for the baryons
which end up
as cold gas or stars in the central $20\ h^{-1}$ kpc at $z=0$.
The arrows show the specific angular momentum of
disk and spheroid stars. 
 }
   \label{j_evolution}
\end{center}
\end{figure}

In Fig.~\ref{j_evolution} we show the evolution of the specific angular momentum
of the dark matter (within the virial radius) and of the cold gas plus stars
(within twice the optical radius) for the no-feedback case (left-hand panel)
and for the feedback case E-0.7 (right-hand panel).
The evolution of the 
specific angular momentum of the dark matter  component
is similar in the two cases, growing as a result of tidal torques
at early epochs and being conserved from turnaround ($z\approx 1.5$)
until $z=0$. On the contrary, the cold baryonic components
in the two cases differ significantly, in
particular at late times. 
In the no-feedback case (NF), 
much angular  momentum is lost through dynamical friction, particularly
through a  satellite which is accreted onto the main halo at $z\sim 1$. 
In E-0.7, on the other hand,  the cold gas
and stars lose rather little specific angular momentum between $z=1$ and $z=0$.
Two main factors contribute to this difference.
Firstly, in E-0.7  a significant number of
young stars form between $z=1$ and $z=0$
with high specific angular momentum (these stars form from
high specific angular momentum gas which becomes cold at late times); and secondly, 
dynamical friction affects the system much less than in NF,
since satellites are less massive.
At $z=0$, disk stars have a specific angular momentum comparable
to that of the dark matter, while spheroid stars have a much
lower specific angular momentum.

\begin{figure}[h]
\begin{center}
 \includegraphics[width=80mm]{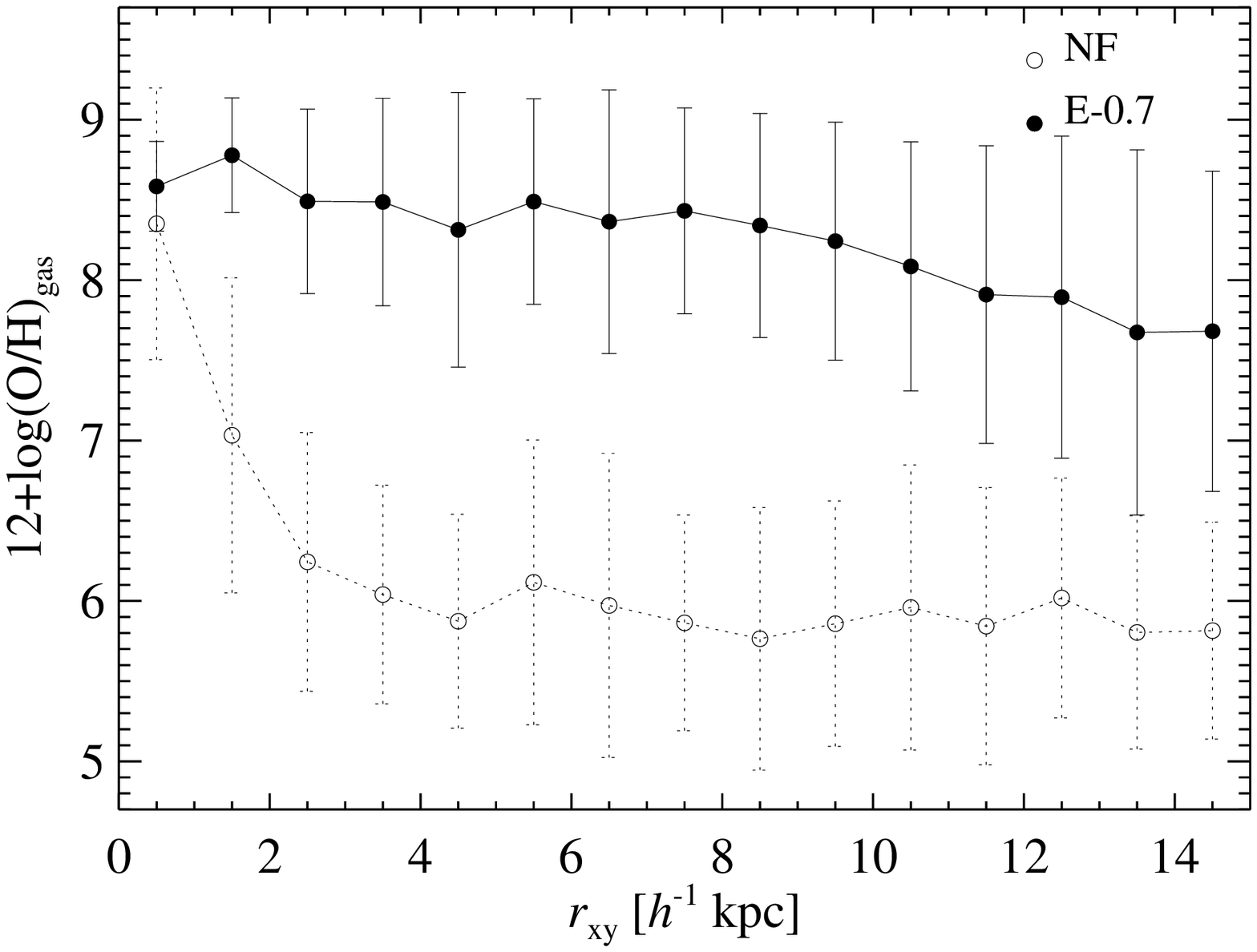}
 \caption{Oxygen abundance for the gas component as a function
of radius projected onto the disk plane for our no-feedback simulation
(NF) and for the feedback case E-0.7. The error bars correspond
to the standard deviation around the mean.
 }
   \label{metal_profiles}
\end{center}
\end{figure}

In Fig~\ref{metal_profiles} we show the oxygen profiles
for the no-feedback (NF) and feedback (E-0.7) runs.
From this figure we can see that SN feedback strongly affects the
chemical distributions. If no feedback is included,
the gas is enriched only in the very central
regions. Including SN feedback triggers a redistribution
of mass and metals through galactic winds and fountains,
giving the gas component a much higher level of enrichment
out to large radii. A linear fit to this metallicity
profile gives a slope of $-0.048$ dex kpc$^{-1}$ and a zero-point
of $8.77$ dex, consistent with the observed values in real
disk galaxies (e.g. Zaritsky et al. 1994).

\begin{figure}[h]
\begin{center}
 \includegraphics[width=80mm]{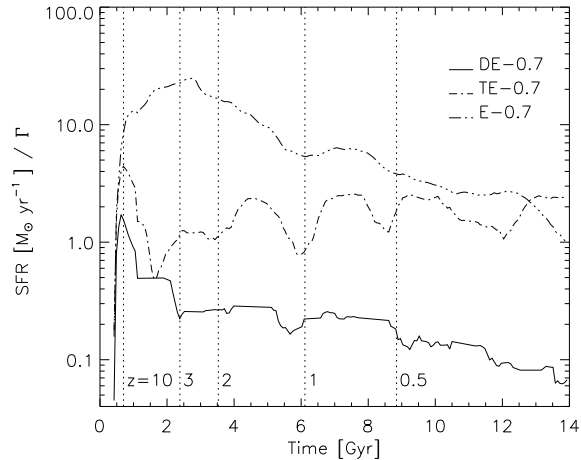}
 \caption{SFRs for simulations  DE-0.7 ($10^{9}\ h^{-1}$ M$_\odot$), 
TE-0.7 ($10^{10}\ h^{-1}$ M$_\odot$) and E-0.7
($10^{12}\ h^{-1}$ M$_\odot$) 
run with energy feedback. To facilitate comparison, the SFRs are
normalized to the scale factor $\Gamma$.
 }
   \label{dwarf}
\end{center}
\end{figure}

Finally, we investigate the effects of SN feedback on different
mass systems. For that purpose we have scaled down our initial
conditions to generate galaxies of $10^{10}\ h^{-1}\ M_\odot$
and  $10^9\ h^{-1}\ M_\odot$ halo mass, and simulate
their evolution including the SN feedback model (with the same
parameters than E-0.7). These simulations are TE-0.7 and
DE-0.7, respectively.
In Fig.~\ref{dwarf} we show the SFRs for these simulations,
as well as for E-0.7, normalized to the scale factor
($\Gamma=1$ for E-0.7, $\Gamma=10^{-2}$ for TE-0.7 and
$\Gamma=10^{-3}$ for DE-0.7).
 From this figure it is clear that SN feedback has
a dramatic effect on small galaxies.
This is because 
more violent winds develop and  baryons are unable to condensate
and form stars. In the smallest galaxy, the SFR is very low at all
times because most of the gas has been lost after the first
starburst episode. This proves that our model is able to reproduce
the expected dependence of SN feedback on virial mass, without
changing the relevant physical parameters.

\section{Conclusions}\label{conclusions}

We have run simulations of a Milky Way-type galaxy
in its cosmological setting in order to investigate the effects of SN feedback
on the formation of galaxy disks. We compare two simulations
with the only difference being the inclusion of the SN energy
feedback model of Scannapieco
et al. (2005, 2006).
Our main results can be summarized as follows:

\begin{itemize}
\item{
SN feedback helps to settle a self-regulated
cycle for star formation  in galaxies, through the
heating and disruption of cold gas and the generation
of galactic winds. 
The regulation of star formation  allows gas to be mantained in a hot
halo which can condensate at late times, becoming a reservoir for
recent star formation. This contributes significantly to the formation
of disk components.
}
\item{When SN feedback is included, the specific angular momentum
of the baryons is  conserved and disks with the correct
scale-lengths are obtained. This results from the
late collapse of gas with high angular momentum, which
becomes available to form stars at later times, when the system does not
suffer from strong interactions.
}
\item{
The injection of SN energy into the interstellar medium generates
a redistribution of chemical elements in galaxies. If energy
feedback is not considered, only the very central regions
were stars are formed are contaminated. On the contrary,
the inclusion of feedback triggers a redistribution of metals
since gas is heated and expands, contaminating the outer
regions of galaxies. In this case, metallicity profiles
in agreement with observations are produced.
}
\item{
Our model is able to reproduce
the expected dependence of SN feedback on virial mass:
as we go to less massive systems, SN feedback has stronger
effects: the star formation rates (normalized to mass) are lower,
and more violent winds develop. This proves that our model
is well suited for studying the cosmological growth of structure
where large systems are assembled through mergers of smaller substructures
and systems form simultaneously over a wide range of scales.
}
\end{itemize}


\end{document}